\documentclass[aps,prb,twocolumn]{revtex4}
\usepackage{graphics,epsfig}

\begin{document}

\title{Quantum Theory:  Interpretation Cannot be Avoided}
\author{Eric Dennis}
\affiliation{Goldman, Sachs \& Co., 85 Broad St., New York, NY 10004 \\
eric.dennis@earthlink.net}
\author{Travis Norsen}
\affiliation{Marlboro College, Marlboro, VT 05344 \\ norsen@marlboro.edu \\ }

\date{\today}

\begin{abstract}
This essay is a response to the (March 2000) Physics Today Opinion 
article ``Quantum Theory Needs No Interpretation'' by Fuchs and Peres. 
It was written several years ago and has been collecting electronic
dust ever since Physics Today said they weren't interested in it.
We post it here with the hope that it may still be of some interest.
\end{abstract}

\maketitle

\begin{quote}
\textit{
In recent years the debate on these ideas has reopened, and there are some
who question what they call ``the Copenhagen interpretation of quantum
mechanics'' -- as if there existed more than one possible interpretation 
of the theory.}  ---Rudolf Peierls\footnote{R. Peierls, \emph{Surprises 
in Theoretical Physics}, Princeton University Press, 1979}
\end{quote}

Some such remark is the first blow of a one-two combination that often
constitutes the entirety of what physics students are taught about the
foundations of quantum mechanics.  In case any recalcitrant youngster
happens to inquire about one of the verboten interpretations, there is
always the second blow: \emph{That} interpretation is merely philosophical 
bias, and therefore no part of \emph{physics}.  Indeed, Peierls' above 
remark occurs in a section which begins with his account of reactionary 
philosophical objections to relativity and quantum theory.

Peierls reiterates this idea in a sequel volume: ``I have discussed the 
interpretation of quantum mechanics and emphasized that there is no 
alternative to the standard view that the wave function (or more generally 
the density matrix) represents our knowledge of a physical system.''\footnote{
R. Peierls, \emph{More Surprises in Theoretical Physics},
Princeton University Press, 1991}

This is surely a commonplace characterization of quantum mechanics.  But 
we should not gloss over how radical a philosophic statement it is:  the
central dynamical object of the theory, the wave function, is said to refer
exclusively to a human mind and \emph{not} any physical system external to 
that mind.

Our objection conerns not primarily the content of this view, although we
strongly disagree with it.  The problem is instead the rather preposterous
notion that this interpretation is uniquely unburdened by any prior
philosophical world-view.

We believe this same inconsistency lies at the core of the Opinion article
authored by Chris Fuchs and Asher Peres.\footnote{C.A. Fuchs and A. Peres, 
``Quantum Theory Needs No `Interpretation' '', Physics Today, 
March 2000, pages 70-71}
They present what they claim is an uncontroversial, bare-bones 
``interpretation without interpretation'' of quantum theory that, they 
assert, is perfectly consistent and free of the extraneous metaphysical 
concepts other interpretations attach to the theory.  They write, for 
example, that ``encumbering quantum mechanics with hidden variables, 
multiple worlds, consistency rules, or spontaneous collapse, without any 
improvement in its predictive power, only gives the illusion of a better 
understanding.''

Yet, like Peierls, what Fuchs and Peres (FP) present as an alternative 
is simply a different but equally philosophic view involving unsupported
metaphysical -- indeed, fundamentally anti-scientific -- assumptions.

\begin{center} ----- \end{center}

In their essay, FP state repeatedly that quantum theory is internally
consistent; this is the reason they dismiss all attempts to ``encumber''
quantum theory with a physical interpretation.  ``[T]o make quantum
mechanics a useful guide to the phenomena around us,'' they write, ``we
need nothing more than the fully consistent theory we already have.  
Quantum theory needs no `interpretation'.''

But is quantum theory internally consistent?  To all appearances, the
answer is:  No.  The theory contains, side by side, two mutually
inconsistent recipes for the time evolution of the wave function:  normal
unitary time evolution according to the Schr\"odinger equation, and the
wave function collapse process which is supposed to occur when a
measurement is made.  The apparent inconsistency between these rules
is the major manifestation of a long-standing difficulty called
the measurement problem.

Resolving, removing, or explaining it away is part of providing
an interpretation of quantum theory.  This typically involves 
clarifying the exact meaning of ``measurement'' by giving some account 
of the primitive ontology of the theory:  which objects in the formalism 
represent objective physical entities and properties, and which represent 
merely our (perhaps limited) knowledge thereof.

Why is the measurement problem a problem at all?  Why not just say
that particles act differently depending on whether or not a measurement
is being made?  This would certainly resolve the contradiction (by showing
that the two rules don't apply at the same time) but it comes at a great
cost:  the ``unprofessionally vague and ambiguous'' \footnote{J. S. Bell, 
\emph{Speakable and Unspeakable in Quantum Mechanics}, Cambridge University 
Press, 1993}
concept of measurement would then play a fundamental role in physics, as 
if ``being observed'' and ``not being observed'' were a physical distinction 
on the same level as, say, the distinction between fermions and bosons.  But 
what precisely is a measurement and how is it distinguished from a
non-measurement?  Moreoever, how could nature possibly care enough about
this human activity to stop her unitary evolution in deference to it?

Something more is therefore needed.  If we don't think that concepts naming 
uniquely human activities belong in our fundamental physical theories, then 
we had better look elsewhere for a resolution of the measurement problem.  
Many different proposals have been made, including the many worlds and 
consistent histories interpretations alluded to by FP.  We will focus, 
however, on Bohm's hidden variable theory \footnote{D. Bohm, Phys. Rev. 
\textbf{85}, 166, 1952; Phys. Rev. \textbf{85}, 180, 1952}
as an example of the theoretical cash value of a realist interpretation 
of quantum mechanics.  We will then compare Bohm's theory to the 
``interpretation without interpretation'' offered by FP.

In any hidden variable theory, the measurement problem is resolved by
supplementing the QM description of the system's physical state (the
wave function) with additional variables describing objective \emph{facts} 
not contained in the wave function itself.  For example, the deBroglie-Bohm 
(dBB) theory supplements the standard quantum description of a particle's
state with a description of the actual position of the particle within
the wave.

Both wave and particle are regarded as real, physical objects.  The 
wave function evolves in time according to Schr\"odinger's equation, while 
the particle's trajectory is determined by the dynamical law
\begin{equation} \label{v}
\vec{v} = 
\frac{\hbar}{m} \mathrm{Im} \left\{ \frac{\vec{\nabla}\psi}{\psi} \right\}
\end{equation}
where $\vec{v}$ is the velocity of the particle.   (The generalization to an
$N$-particle state and/or to particles with spin is straightforward and 
natural.)  
Thus the point-like scintillations seen 
in the double-slit experiment, and measurement results generally, represent 
the positions of the actual particles themselves, and Equation (\ref{v}) 
can be shown \footnote{Ibid.} to imply consistency between the 
probability distribution associated with these particles and that defined 
by $|\psi|^2$.

Granting real, physical existence to at least some objects of the theory in 
this way has a number of positive effects. Primarily, it solves the measurement
problem quantitatively and unambiguously.  There is no longer any mystery
of how measurements end with definite results.  Thus, there is no 
longer any need for a separate
postulate about the collapse of the wave function when a ``measurement''
occurs. Rather, a theorem is derived regarding the \emph{effectively}
irreversible behavior of highly entangled wave functions associated with
the introduction of many additional degrees of freedom describing a
measurement apparatus (which, in dBB, is simply another ordinary physical
object in the universe).

This works because the wave function need be considered only insofar as it
affects the particles. If the wave function splits into a number of
distinct and independently evolving packets in configuration space (which
is guaranteed under certain well-defined conditions by the
decoherence effect), then the particles are influenced only by the one
packet in which they actually reside. The particles behave as if all the
other packets have been collapsed away, but without any need to postulate
a non-unitary, Schr\"odinger-violating collapse process.

An additional attractive aspect of dBB is that these particles can now be
viewed as primary physical objects, just like in classical physics, giving
us an intuitive physical picture of even the most complicated many body
quantum processes.  This is in fact a good thing, despite the
abundant philosophical antagonism directed against it over the last
century.

The dBB theory is thus a major step toward professionalism and
precision in the foundations of quantum mechanics. Fuchs and Peres,
however, are quick to dismiss dBB and the other interpretations
mentioned above on the grounds that they are unnecessarily encumbered
with philosophical concepts.  What do they offer in place of these 
interpretations?

They offer a series of radical philosophic statements regarding the
meaning of the wave function: it is ``only a mathematical expression for
evaluating probabilities''; contradictions arise from ``attributing
reality to quantum states''; ``no wavefunction exists either before or
after we conduct an experiment''; ``the wavefunction is not an objective
entity''; ``collapse is something that happens in our description of the
system, not to the system itself''; ``the time dependence of the
wavefunction does not represent the evolution of a physical system.''

The idea of the wave function as nothing but a means of computing
probabilities immediately raises the question:  probabilities \emph{of
what}?  FP do not believe that these are probabilities of real subatomic 
events, \emph{e.g.}, the standard proposition that $|\psi(x)|^2$ represents 
the probability of detecting a particle to \emph{be} at $x$.  That would 
be a reversion to micro-physical realism, in which an electron is assumed 
capable of \emph{being} at all -- just the kind of thing FP pointedly dismiss 
as a wrong-headed attempt to ``create a new theory with features that 
correspond to some reality independent of our potential experiments.''

Instead, FP's answer is that the probabilities calculated in quantum
theory refer only to macroscopic events.  But this simply places a
different name on the same ambiguity that had previously been shuffled
under the idea of ``measurement.'' Where exactly is the cut between micro-
and macroscopic, and why should such a cut enter into the fundamental laws
of physics?  If one electron is not objectively real, and two electrons
are not objectively real, why should a collection of $10^{23}$ electrons
be real?  Things like temperature and elasticity may be emergent
properties, but surely \emph{existence as such} is not.  Nothing real can
emerge from that which doesn't exist.  Like the proposed divide
between measurement and non-measurement, an artificial cut between micro-
and macro-worlds simply does not belong in the fundamental laws of nature.

To FP, however, this is not a problem, because they hold a gravely
epistemic view of the quantum world.  For them, the wave function -- which
is to say the entire content of conventional quantum theory -- refers to
something mental, not something in external reality.  They do attempt to
provide a hedge by admitting that ``the possible existence of an objective
reality independent of what observers perceive'' cannot be excluded in
principle.  Still, even if there is an objective reality, they argue, it
is unimportant in practice for ``using the theory and understanding its
nature.''

This sort of radical anti-realism can resolve interpretative paradoxes
in virtually any context, and it has been attempted before. For example,
Mach's rejection of the need to ground ``pressure'' and ``temperature'' in
terms of real microscopic entities obviates the subsequent puzzles
associated with thermodynamic equilibrium and the convergence to it.  
This does not, however, appear to be good evidence against the reality of
atoms.  More broadly, the philosophical doctrine of solipsism
can ``solve'' every difficult problem in the history of science by
simply denying that anything but the advocate's own mental experiences
exist.  This, however, is infinitely distant from the kind of solution we 
are interested in as scientists.

\begin{center} ----- \end{center}

Alan Sokal once described the motivation for his famous prank as follows:  
``What concerns me is the proliferation not just of nonsense and sloppy
thinking \emph{per se}, but of a particular kind of nonsense and sloppy
thinking: one that denies the existence of objective realities, or (when
challenged) admits their existence but downplays their practical
relevance.'' \footnote{A. D. Sokal, ``A Physicist Experiments with Cultural 
Studies'', \emph{Lingua Franca}, May/June 1996} 
Sokal was, of course, referring to the depths of irrationality to which 
some fields in the humanities and social sciences have descended.

It is frightening that the same anti-realism Sokal ridiculed there can be
put forward as a supposedly natural and uncontroversial interpretation of
quantum theory.  It is even more frightening that this anti-realism 
is glibly passed off as non-philosophical, for this suggests not only
that many physicists have accepted a fundamentally anti-scientific set 
of philosophical ideas but that, in addition, they have done so unwittingly.

Does the community of \emph{physicists} -- those individuals whose lives
are dedicated to observing, understanding, and learning to control the
physical world -- really accept as solid, hard-nosed science the idea
(hedged or not) that there is no physical world -- that, instead, it's all
in our minds?

\end{document}